\documentclass{ws-procs961x669}            
\begin{document}
\title{Highly magnetized white dwarfs: implications and current status}

\author{B. Mukhopadhyay$^{1*}$, M. Bhattacharya$^{2}$, A. J. Hackett$^{3}$, S. Kalita$^{1}$,\\ 
D. Karinkuzhi$^{1}$ and C. A. Tout$^{3}$}

\address{${}^1$Department of Physics, Indian Institute of Science, Bangalore 560012, India\\
$^*$E-mail: bm@iisc.ac.in\\
${}^2$Department of Physics, The Pennsylvania State University, University Park, PA 16802, USA\\
${}^3$Institute of Astronomy, University of Cambridge, Madingley Road, Cambridge CB3 0HA, UK}

\begin{abstract}
Over the last decade or so, we have been developing the possible existence of highly magnetized white dwarfs with analytical stellar structure models. While the primary aim was to explain the nature of the peculiar overluminous type Ia supernovae, later on, these magnetized stars were found to have even wider ranging implications including those for white dwarf pulsars, soft gamma-ray repeaters and anomalous X-ray pulsars, as well as gravitational radiation. In particular, we have explored in detail the mass-radius relations for these magnetized stars and showed that they can be significantly different from the Chandrasekhar predictions which essentially leads to a new super-Chandrasekhar mass-limit. Recently, using the stellar evolution code STARS, we have successfully modelled their formation and cooling evolution directly from the magnetized main sequence progenitor stars. Here we briefly discuss all these findings and conclude with their current status in the scientific community.
\end{abstract}


\bodymatter

\section{Introduction}
\label{sec1}
More than a dozen overluminous Type Ia supernovae (SNe Ia) have already been observed \cite{Howell06,Scalzo10} which strongly suggest the existence of massive progenitors with masses $M > 2M_{\odot}$. Although the evolutions of accreting or rapidly differentially rotating white dwarf (WD) binaries have been used to explain such progenitors \cite{Hachisu86,YL04}, these models are unable to explain masses up to $2.8M_{\odot}$ that are inferred from the observations. Highly magnetised super-Chandrasekhar WDs (B-WDs) have been recently proposed as the possible progenitors of these peculiar overluminous SNe Ia. In addition to SNe Ia, B-WDs are also considered as promising candidates for soft gamma-ray repeaters (SGRs) and anomalous X-ray pulsars (AXPs) at lower magnetic fields than neutron star (NS) based magnetar, 
satisfying their ultraviolet luminosity cut-off \cite{MR16}. 

It has been shown that strong magnetic fields can modify the equation of state (EoS) of electron degenerate matter and yield super-Chandrasekhar WDs \cite{DM12,DM13,SM15}, irrespective of the rotation rate. Indeed, the observational data from the Sloan Digital Sky Survey (SDSS) suggest that magnetized WDs tend to have larger masses than their non-magnetic counterparts, even though they span the same effective temperature range \cite{Vanlandingham05,Ferrario15}. Inspired by above
findings, the effect of strong magnetic fields on the stellar structure, for various field configurations, has been explored for both Newtonian \cite{DM12} and general relativistic formalisms \cite{DM14,DM15,SM15}.

Magnetized WDs have many important implications apart from their link to peculiar SNe Ia and hence their other properties are worth exploring \cite{MR16,BM17a,BM17b}. Recent works \cite{MB18,MB_euro18,G20,MB21} have shown that B-WDs can be too dim to detect directly, and have also explored \cite{KM19,Kalita20} the ability of rotating B-WDs to generate gravitational radiation which can be detected by the space-based gravitational wave detectors. Furthermore, other additional physics such as modified gravity \cite{Ban17,EPL19}, ungravity effect \cite{BertM16}, effects of net charge \cite{Liu14,Car18}, lepton number violation \cite{Belyaev15} and anisotropic pressure \cite{HB13} have also been explored to show the possible
existence of super-Chandrasekhar WDs. Here we discuss the broad implications of these magnetized stars as well as their current status.

\section{Origin and evolution of strong magnetic fields}
It has been well known that purely poloidally or toroidally dominated fields are both structurally unstable \cite{MT73,Tayler73}. 
However, it also has been shown that magnetized WDs with toroidally dominated mixed field configuration (along with small poloidal component) are one of the most plausible cases \cite{Wick13} and have approximately spherical shape \cite{SM15}. Although the surface fields can be observationally inferred, the interior field cannot be directly constrained. 
Nevertheless, there is a sufficient evidence that the stars exhibit dipolar fields in their outer regions and hence are expected to have stronger interior (toroidally dominated) fields than at the surface. Numerical simulations have indeed shown that central fields of B-WDs can be several orders of magnitude higher than the surface field \cite{SM15,BM17a,QT18}.

\begin{figure}[h]%
\begin{center}
  \parbox{2.4in}{\includegraphics[width=2.4in]{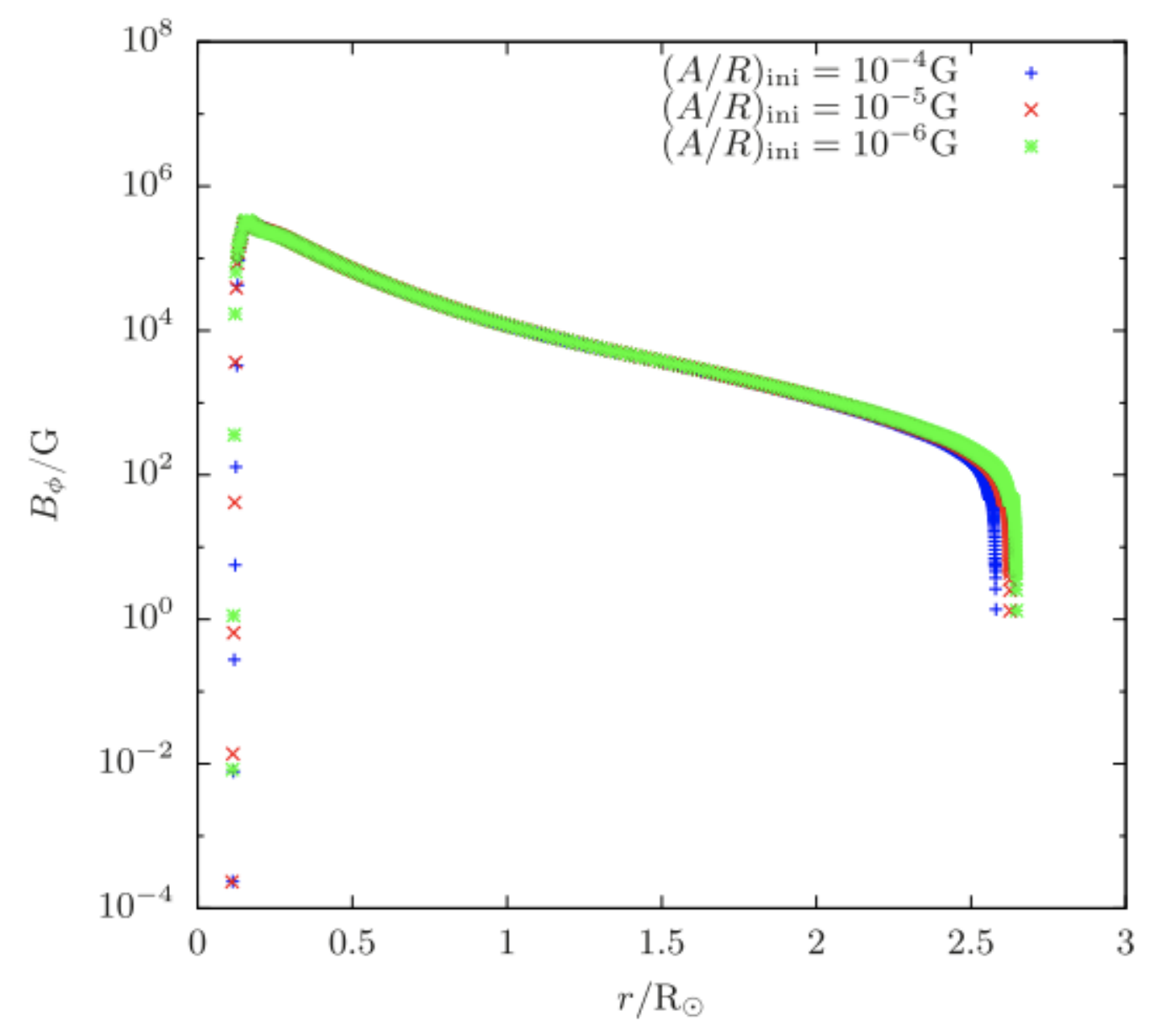}}
  \hspace*{1pt}
  \parbox{2.4in}{\includegraphics[width=2.4in]{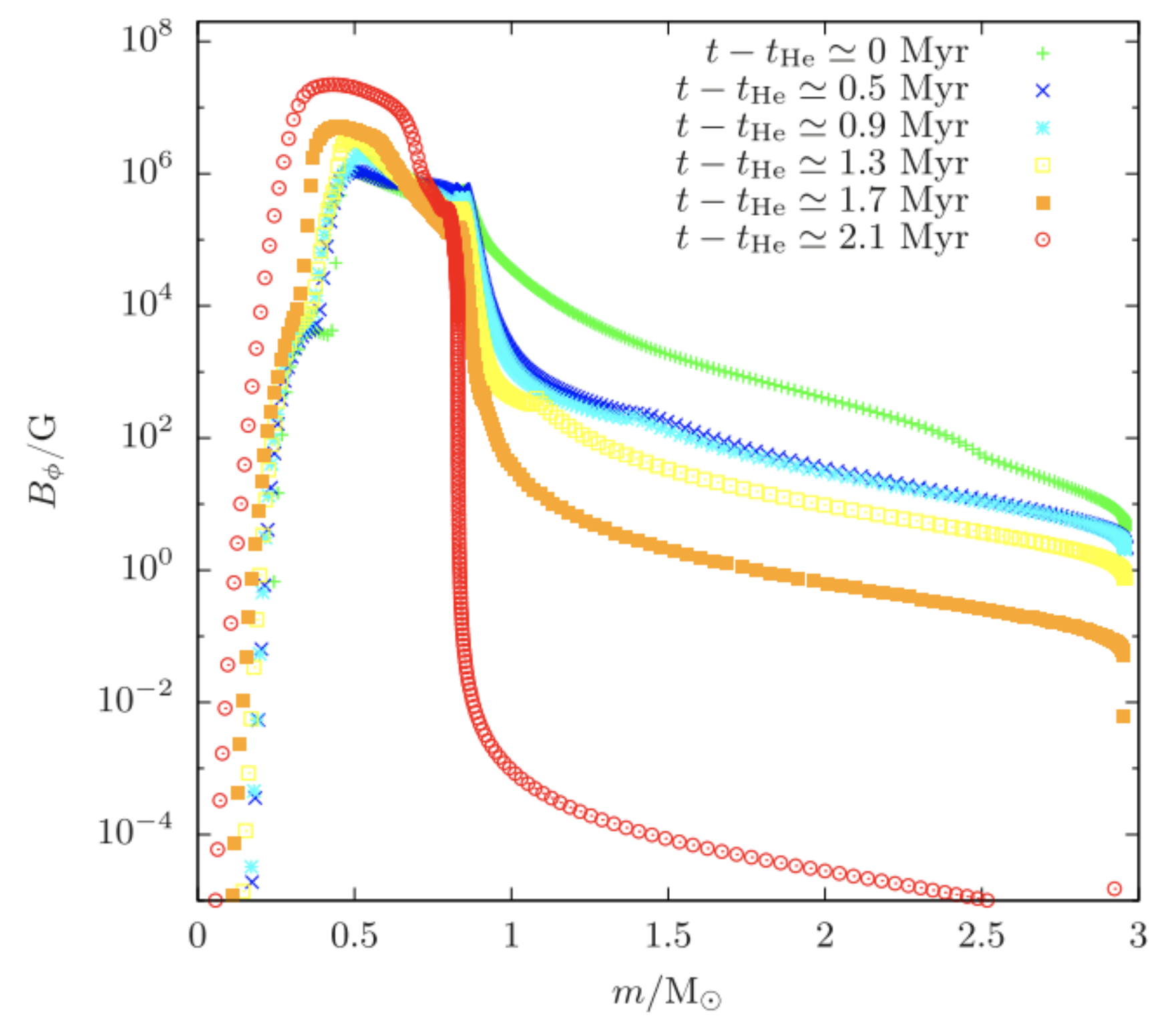}}
  \caption{ 
	{\it Left panel:} Toroidal field in the interior of the star as a function of the radius at the end of the main sequence.
	{\it Right panel:} Toroidal field inside the star as a function of the mass coordinate at various times after 
	the helium exhaustion in the core, during the asymptotic giant phase.}%
 \label{fig1}
\end{center}
\end{figure}

The evolution of the poloidal and toroidal magnetic field components along with the angular momentum has been modelled recently with the Cambridge stellar evolution code STARS \cite{QT18}, using advection-diffusion equations coupled to the structural and compositional equations of stars. They have shown that the magnetic field is likely to be dipolar, decaying as an inverse square law for most of the star. Their results also suggest that at the end of main sequence, the star may have toroidally dominated magnetic fields. The left panel of Figure \ref{fig1} shows the evolution of toroidal field in the stellar interior as a function of radius at the end of main sequence, while the right panel shows the field as a function of the mass coordinate at various times after the helium exhaustion in the core, during the asymptotic giant phase. 

Large-scale magnetic fields can be governed in the degenerate core of B-WDs even during the late stages of stellar evolution \cite{QT18}, and very high fields can develop in these stars based on the conservation of magnetic flux, besides 
from the dynamo mechanism. Hence, strong fields inside magnetized WDs can also be of fossil origin \cite{DMR13}. 
This can be understood as follows. While the mass of the WD increases due to accretion, magnetic field is advected into its interior. Consequently, the gravitational power dominates over the degeneracy pressure leading to the contraction of the star. The initial seed magnetic field is then amplified as the total magnetic flux remains conserved. For magnetic field $B \sim 10^8\, {\rm G}$ in a star of size $R \sim 10^6\, {\rm km}$, the resultant flux will be $\sim 10^{20}\, {\rm G\, km^2}$. From flux freezing, for a 1000 km size B-WD, the magnetic field can then grow up to $\sim 10^{14}\, {\rm G}$. Once the field increases, the total outward force further builds up to balance the inward gravitational force and the whole cycle is repeated multiple times. Therefore, the magnetic fields of highly magnetized WDs are likely to be fossil remnants from their main-sequence progenitor stars.

\begin{figure}[h]%
\begin{center}
  \parbox{2.4in}{\includegraphics[width=2.4in]{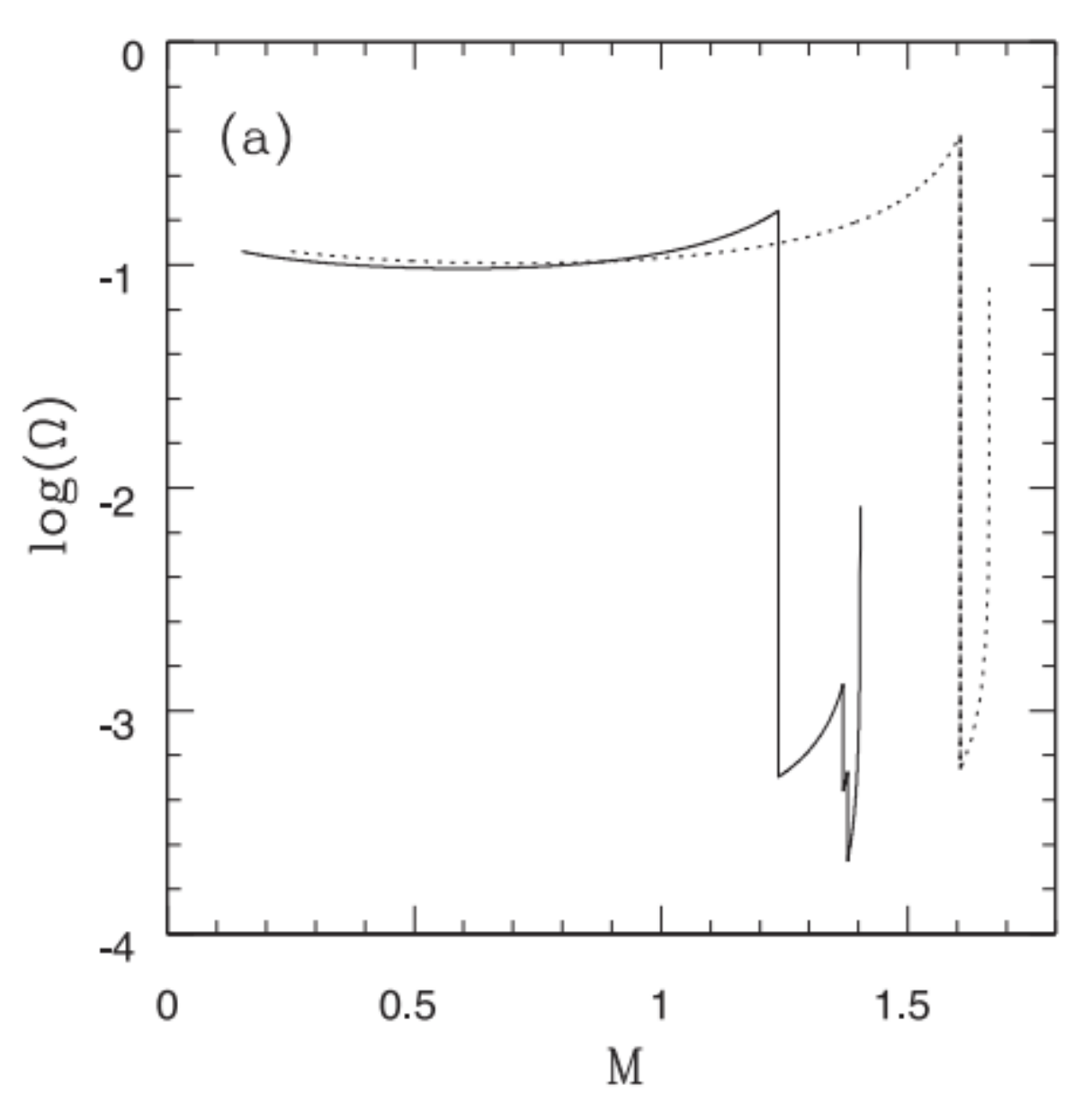}}
  \hspace*{1pt}
  \parbox{2.4in}{\includegraphics[width=2.4in]{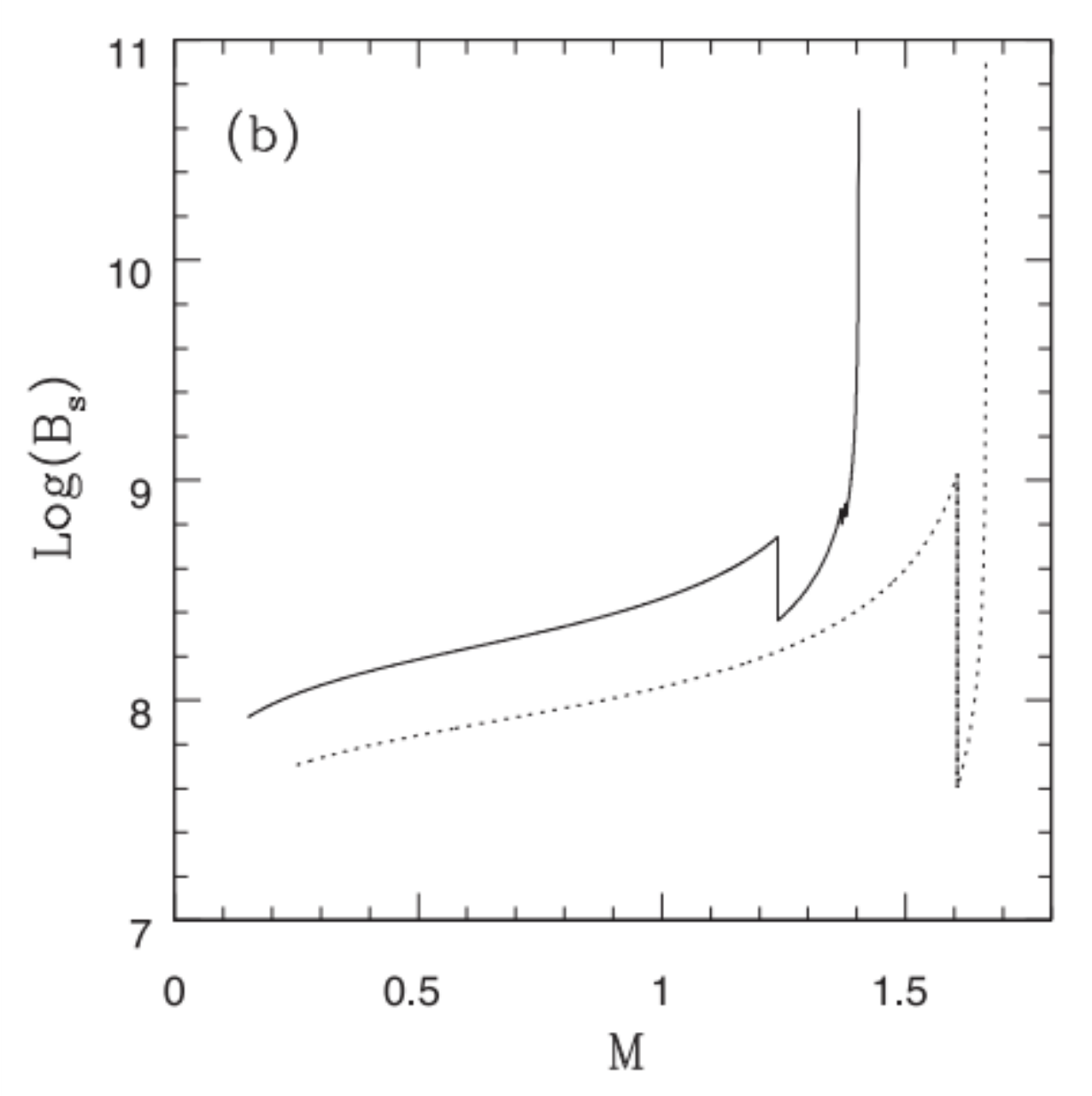}}
  \caption{Time evolution of angular velocity (left) and magnetic field (right) as functions of mass. The solid curves correspond to the case with $n=3$, $m=2.7$, $\rho=0.05\, {\rm g\, cm^{-3}}$, $l=1.5$ and dotted curves correspond to the case with $n=3$, $m=2$, $\rho=0.1\, {\rm g\, cm^{-3}}$, $l=2.5$. The other parameters are fixed with $k=10^{-14}$, $\dot{M}=10^{-8}\, M_{\odot} {\rm yr}^{-1}$, $\alpha=10^{\circ}$ and $R=10^4\, {\rm km}$ at $t=0$.}%
 \label{fig2}
\end{center}
\end{figure}

Repeated episodes of accretion and spin-down have also been suggested as a plausible mechanism that can lead to a highly magnetized WD \cite{BM17a}. Here, the entire evolution of the B-WD can be classified in two phases: accretion-powered and rotation-powered. The accretion-powered phase is governed by three conservation laws: linear and angular momenta conservation and conservation of magnetic flux, around the stellar surface, given by
\begin{eqnarray}
l\Omega(t)^2 R(t) = \frac{GM(t)}{R(t)^2}, \nonumber\\
I(t)\Omega(t) = {\rm constant}, \nonumber \\
B_s(t)R(t)^2 = {\rm constant},
\end{eqnarray}
where $l$ accounts for the dominance of gravitational force over the centrifugal force, hence $l>1$, $I$ is the stellar moment of inertia and $\Omega$ is the angular velocity of the star that includes contribution acquired due to accretion as well. Solving these equations simultaneously gives the time evolution of radius, magnetic field and angular velocity during the accretion phase. Accretion discontinues when
\begin{equation}
-\frac{GM}{R^2} = \frac{1}{\rho} \frac{d}{dr} \left(\frac{B^2}{8\pi}\right)_{r=R} \sim -\frac{B_s^2}{8\pi R\rho},
\end{equation}
where $\rho$ is the density of the inner disk edge. If the magnetic field is dipolar, $\dot{\Omega} \propto \Omega^3$ for a fixed magnetic field. Generalizing it to $\dot{\Omega} = k\Omega^n$ with constant $k$ giving
for the spin-powered phase, we obtain
\begin{eqnarray}
	\Omega &=& [\Omega_0^{1-n} - k(1-n)(t-t_0)]^{1/1-n}, \\
	{\rm and}\,\,\,B_s &=& \sqrt{\frac{5c^3 I k \Omega^{n-m}}{R^6 {\rm sin}^2 \alpha}}.
\end{eqnarray}
Here $\Omega_0$ is the initial angular velocity for the spin-powered phase (once accretion stops) at time $t=t_0$. The value of $k$ is fixed such that $B_s$ can be constrained at $t=t_0$, which is known from the field evolution in the preceding accretion-powered phase. Here $n=m=3$ corresponds to the dipole field configuration, therefore $m$ represents the deviation from dipolar field, especially for $n=3$. Figure \ref{fig2} shows the sample evolutions of angular velocity and magnetic field as functions of stellar mass. In both cases, initially larger $\Omega$ with accretion drops significantly during the spin-powered phase, followed by a phase of its increasing trend. At the end of the evolution, the star can be left either as a super-Chandrasekhar WD and/or an SGR/AXP candidate with a higher spin frequency. 

\section{Rotating magnetized white dwarfs}
\begin{figure}[h]%
\begin{center}
  \parbox{1.6in}{\includegraphics[width=1.6in]{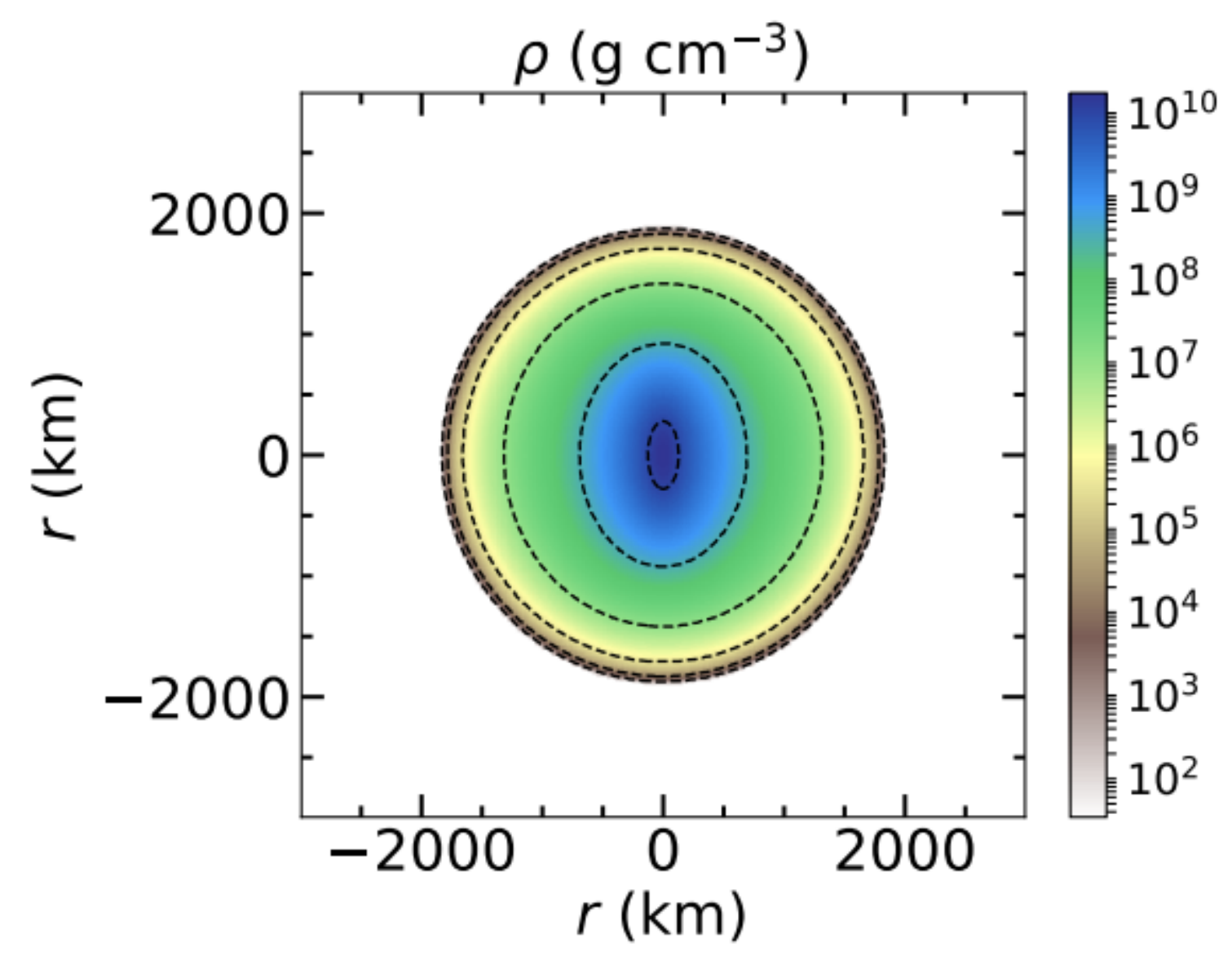}}
  \hspace*{0.25pt}
  \parbox{1.6in}{\includegraphics[width=1.6in]{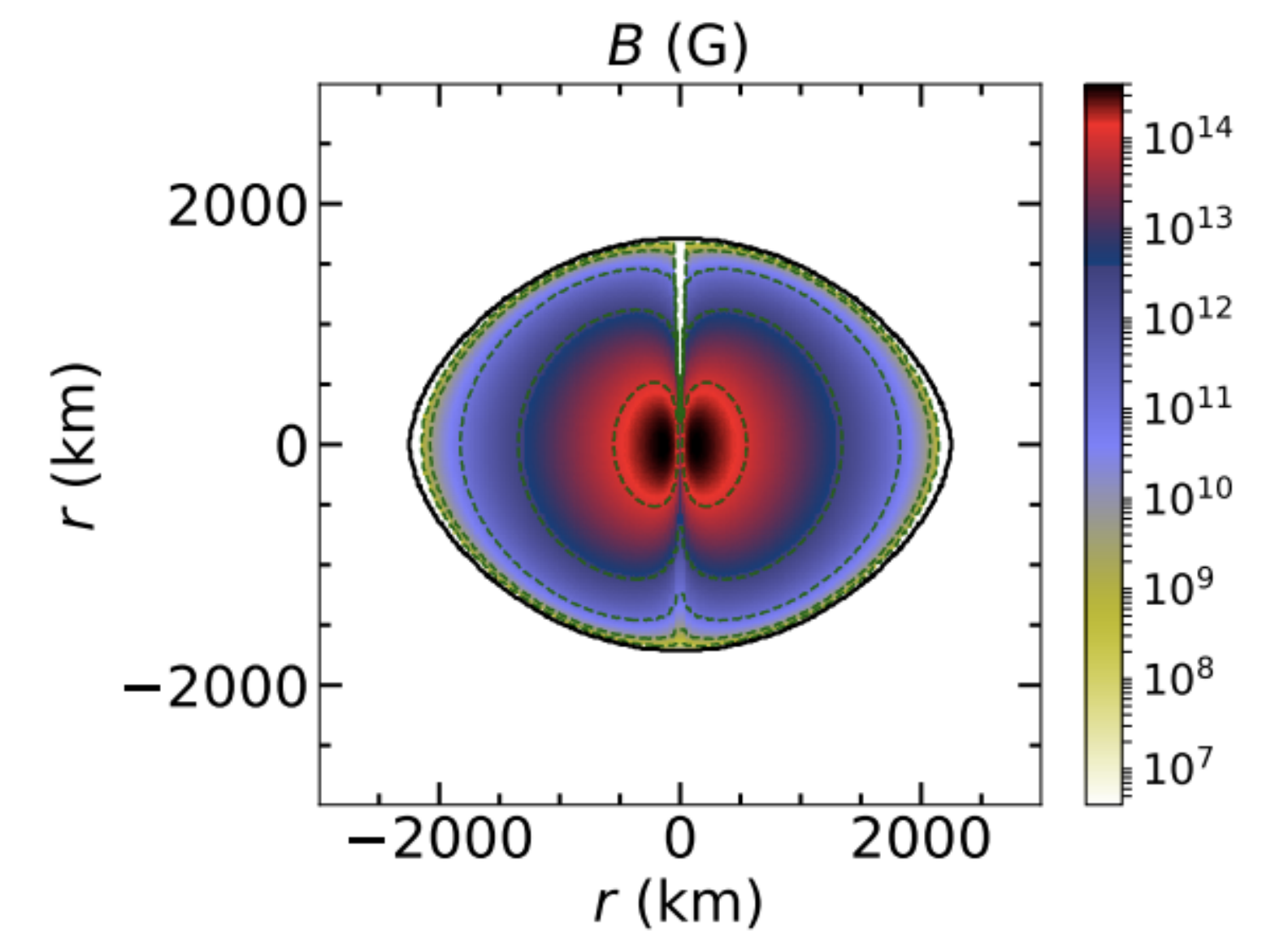}}
  \hspace*{0.25pt}
  \parbox{1.6in}{\includegraphics[width=1.6in]{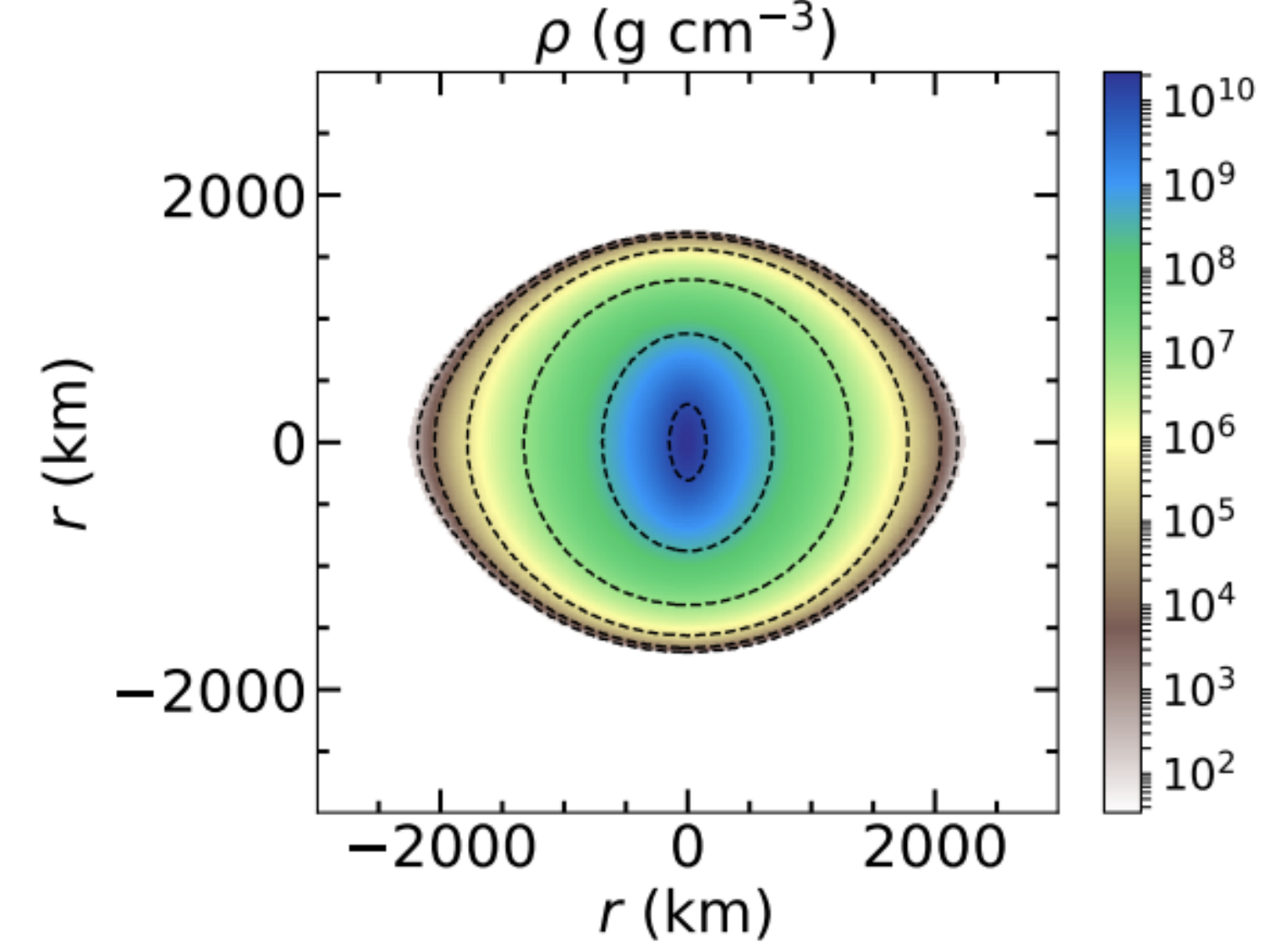}}
  \caption{Density and magnetic field strength isocontours for uniformly rotating B-WDs with toroidal magnetic field are shown in the left and center panels, respectively. The parameters used for both are $\Omega \sim 0.0628\, {\rm rad/s}$, $B_{max} \sim 2.7\times10^{14}\, {\rm G}$, ME/GE $\sim 0.1$, KE/GE $\sim 3.6\times10^{-6}$. The density isocontours for the $\Omega \sim 3.6537\, {\rm rad/s}$, $B_{max} \sim 2.7\times10^{14}\, {\rm G}$, ME/GE $\sim 0.1$, KE/GE $\sim 1.3\times10^{-2}$ case is shown in the right panel.}%
\label{fig3}
\end{center}
\end{figure}

Although, in nature, WDs are expected to consist of mixed fixed geometry, here we consider toroidally dominated magnetic field cases as they ensure the stability of these stars. It has been shown that toroidally dominated (and purely toroidal) field not only makes the star (slightly) prolate but also increases its equatorial radius \cite{IS04,FR12,SM15}. Figure \ref{fig3} shows specific cases for toroidal field configuration combined with rotation. In the left panel, as the angular frequency is small, it does not affect the star considerably and results in a marginally prolate star. In contrast, the right panel, due to high angular velocity, exhibits that the low density region is affected more due to rotation than the high density region, resulting in an oblate shaped WD. From the magnetic field strength isocontours shown in the center panel, we can see that the surface magnetic field can decrease up to $\sim 10^9\, {\rm G}$ even if the central field is $\sim 10^{14}\, {\rm G}$. For both cases, the magnetic to gravitational energies ratio (ME/GE) as well as 
kinetic to gravitational energies ratio (KE/GE) are chosen to be $\lesssim 0.1$ to maintain stable equilibrium \cite{ChF53,Komatsu89,Braithwaite09}.

\begin{figure}[h]%
\begin{center}
  \parbox{2.4in}{\includegraphics[width=2.4in]{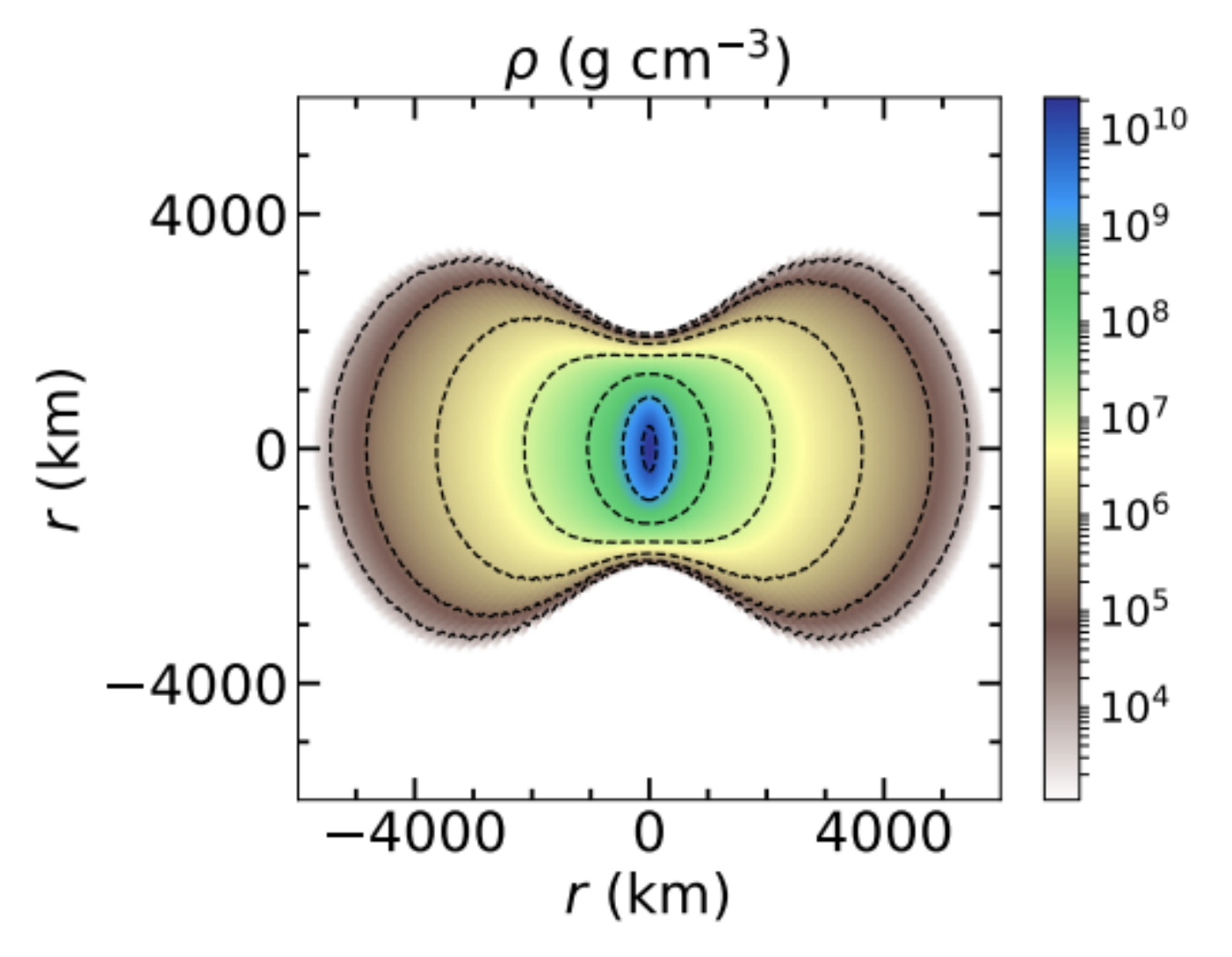}}
  \hspace*{1.0pt}
  \parbox{2.4in}{\includegraphics[width=2.4in]{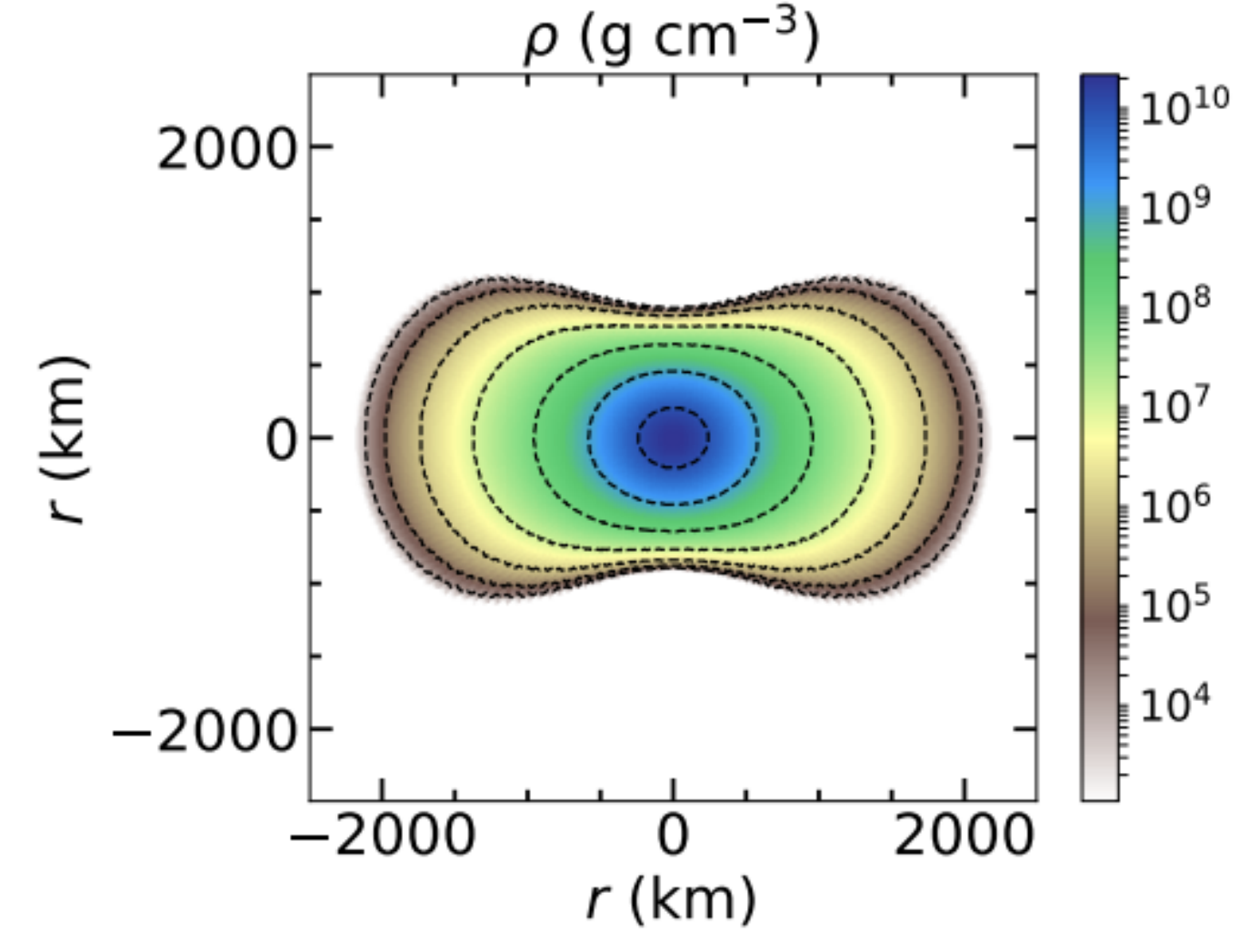}}
  \caption{Density isocontours for differentially rotating magnetized WD is shown. For the left panel, we have toroidal field with $\Omega \sim 0.62\, {\rm rad/s}$, $\Omega_c \sim 10.15\, {\rm rad/s}$, $B_{max} \sim 3.2\times10^{14}\, {\rm G}$, ME/GE $\sim 0.14$, KE/GE $\sim 0.1$. For the right panel, we use poloidal field with $\Omega \sim 2.74\, {\rm rad/s}$, $\Omega_c \sim 20.30\, {\rm rad/s}$, $B_{max} \sim 3.9\times10^{11}\, {\rm G}$, ME/GE $\sim 7.8\times10^{-8}$, KE/GE $\sim 0.14$.}%
\label{fig4}
\end{center}
\end{figure}

We have also considered differentially rotating B-WDs \cite{SM15}. The angular velocity profile in the XNS code is specified as \cite{Ster03,BDZ11}
\begin{equation}
F(\Omega) = A^2 (\Omega_c - \Omega) = \frac{R^2 (\Omega - \omega)}{\alpha^2 - R^2 (\Omega- \omega)^2},
\end{equation}
where $A$ is a constant that indicates the extent of differential rotation, $R = \psi^2 r {\rm sin}\theta$, $\omega = -\beta^{\phi}$, $\Omega_c$ is the angular velocity at the center and $\omega$ is the angular velocity at radius $r$. Figure \ref{fig4} shows the density isocontours of differentially rotating B-WDs for toroidal field with surface 
angular velocity $\Omega \sim 0.62\, {\rm rad/s}$, $\Omega_c \sim 10.15\, {\rm rad/s}$, $B_{max} \sim 3.2\times10^{14}\, {\rm G}$, ME/GE $\sim 0.14$, KE/GE $\sim 0.1$ (left panel) and poloidal field with $\Omega \sim 2.74\, {\rm rad/s}$, $\Omega_c \sim 20.30\, {\rm rad/s}$, $B_{max} \sim 3.9\times10^{11}\, {\rm G}$, ME/GE $\sim 7.8\times10^{-8}$, KE/GE $\sim 0.14$ (right panel). It can be seen that `polar hollow' structure can form with differential rotation regardless of the specific geometry of the magnetic field.

\section{Non-rotating magnetized WDs with finite temperature}
Apart from increasing the limiting mass of WDs, strong magnetic fields can also influence the thermal properties such as luminosity, temperature gradient and cooling rate of the star \cite{MB18,G20,MB21}. In order to model such a WD, the 
total pressure inside the star is modelled by including the contributions from the degenerate electron gas, ideal gas and magnetic pressures. The interface is defined to be the radius where the contributions from electron degenerate core and outer ideal gas pressures are equal. The presence of strong fields gives rise to additional pressure $P_B = B^2/8\pi$ and density $\rho_B = B^2/8\pi c^2$ inside the magnetized WDs \cite{Sinha13}. 

Assuming the B-WD to be approximately spherical, the model equations for magnetostatic equilibrium, photon diffusion and mass conservation can be written within a Newtonian framework as 
\begin{eqnarray} 
&\frac{{\rm d}}{{\rm d}r}(P_{\rm deg}+P_{\rm ig}+P_{B}) = -\frac{Gm(r)}{r^2}(\rho + \rho_B), \nonumber \\
&\frac{{\rm d}T}{{\rm d}r} = -{\rm max}\left[\frac{3}{4ac} \frac{\kappa \rho}{T^3} \frac{L_r}{4\pi r^2}, \left(1 - \frac{1}{\gamma}\right)\frac{T}{P} \frac{{\rm d}P}{{\rm d}r}\right], \vspace{0.1in} \nonumber \\
&\frac{{\rm d}m}{{\rm d}r} = 4\pi r^2 (\rho + \rho_B).
\end{eqnarray}
Here we have ignored the magnetic tension terms for radially varying $B$. In these equations, $P_{\rm deg}$ and $P_{\rm ig} = \rho kT/\mu m_p$ are the electron degeneracy pressure and hence the ideal gas pressure respectively, $\rho$ is the matter density, $T$ is the temperature, $m(r)$ is the mass enclosed within radius $r$, $\kappa$ is the radiative opacity, $L_r$ is the luminosity at radius $r$, and $\gamma$ is the adiabatic index of the gas. 

The opacity in the surface layers of non-magnetised WD is approximated with the Kramers' formula, $\kappa = \kappa_0 \rho T^{-3.5}$, where $\kappa_0 = 4.34\times10^{24}Z(1+X)\ {\rm cm^2\ g^{-1}}$, and $X$ and $Z$ are the mass fractions of hydrogen and heavy elements (other than hydrogen and helium) in the stellar interior respectively. To capture the radial variation of the field magnitude within the B-WD, we adopt a profile used extensively to model magnetized NSs and B-WDs \cite{Band97,DM14}, given by
\begin{equation}
B\left(\frac{\rho}{\rho_0}\right) = B_{\rm s} + B_0 \left[1 - {\rm exp}\left(-\eta \left(\frac{\rho}{\rho_0}\right)^{\gamma}\right)\right],
\label{Bprof}
\end{equation}
where $B_s$ is the surface magnetic field, $B_0$ is a fiducial magnetic field, $\eta=0.8$ and $\gamma=0.9$ are dimensionless parameters along with $\rho_0=10^9\, {\rm g/cm^3}$ determining how the field decays from the core to the surface. The radial luminosity can be assumed to be constant so that $L_r = L$ as there is no hydrogen burning or other nuclear fusion reactions taking place within the WD core. We solve the differential equations by providing the surface density, mass and surface temperature as the boundary conditions.

\begin{figure}[h]%
\begin{center}
  \parbox{2.4in}{\includegraphics[width=2.4in]{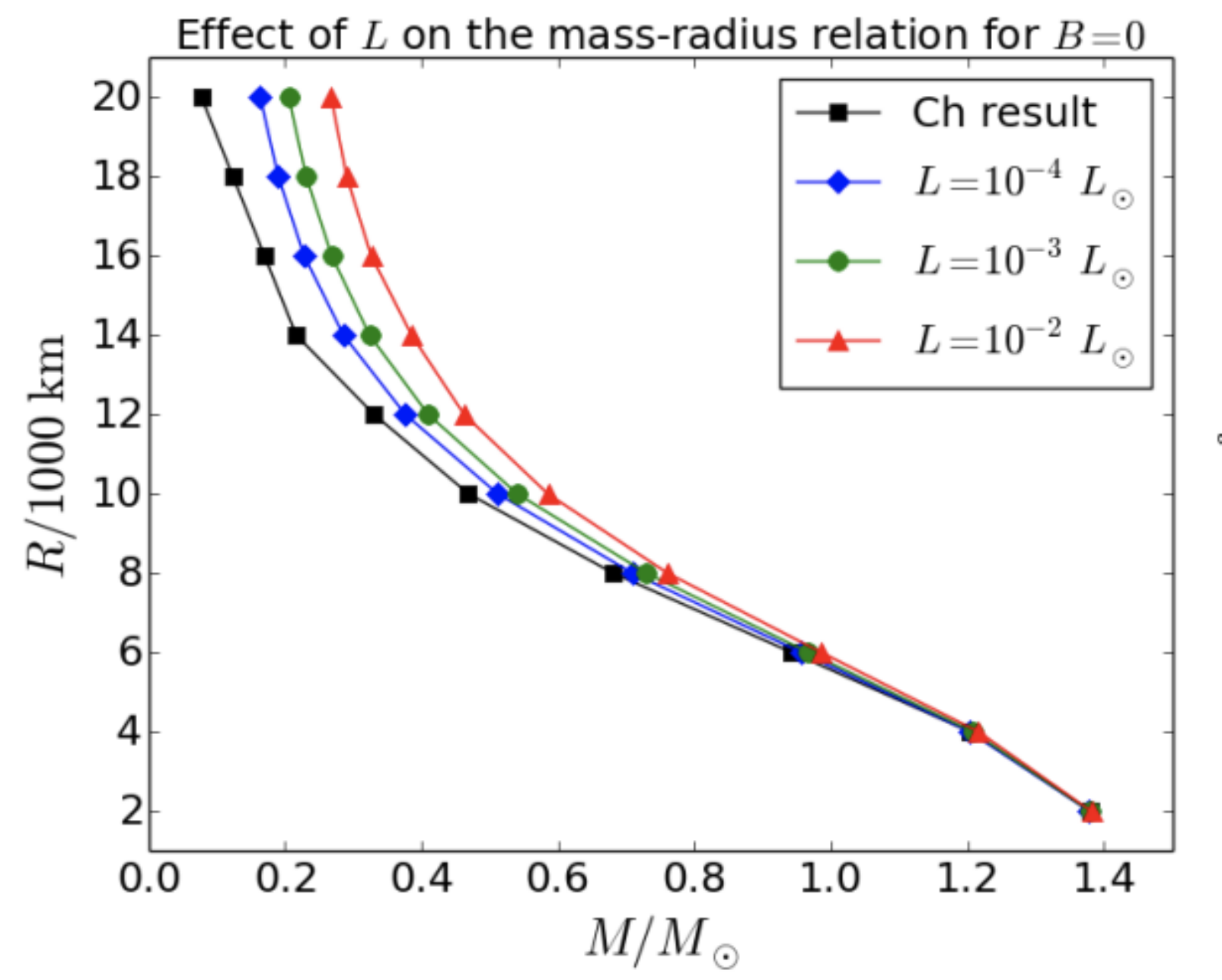}}
  \hspace*{1.0pt}
  \parbox{2.4in}{\includegraphics[width=2.4in]{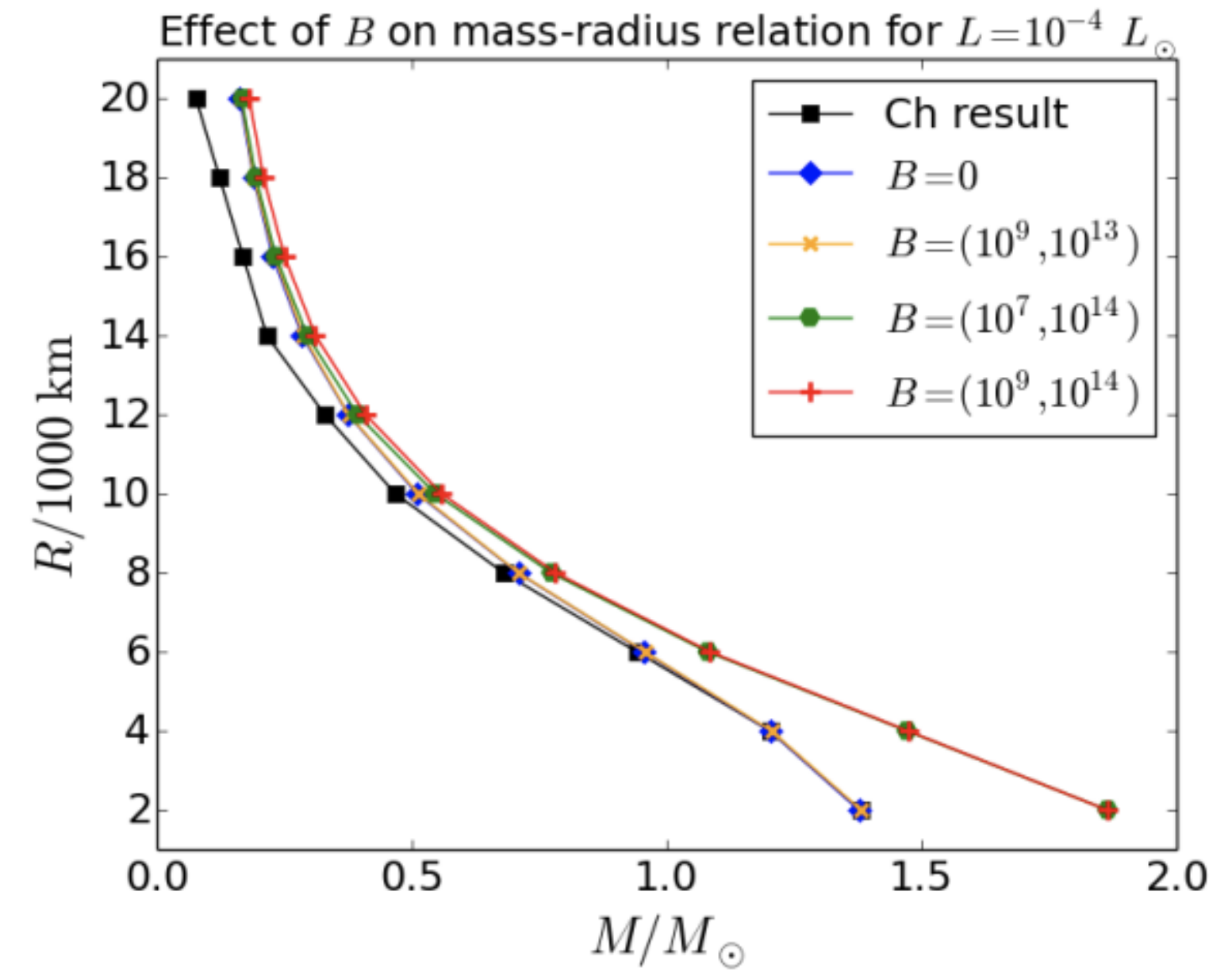}}
  \caption{\emph{Left panel:} The effect of $L$ on the mass--radius relation of non-magnetised WDs is shown for 
	$L=10^{-4}\ L_{\odot}$ (blue diamonds),
$L=10^{-3}\ L_{\odot}$ (green circles) and $L=10^{-2}\ L_{\odot}$ (red triangles), along with the Chandrasekhar result (black squares). \emph{Right panel:} The effect of field strength on the mass--radius relation of B-WDs is shown for 
	$B=(B_s,B_0)=(0,0)$ (blue diamonds), $B=(10^9,10^{13})\ {\rm G}$ (orange crosses), $B=(10^7,10^{14})\ {\rm G}$ (green circles) and $B=(10^9,10^{14})\ {\rm G}$ (red pluses), along with the Chandrasekhar result (black squares) for $L=10^{-4}\ L_{\odot}$.}%
\label{fig5}
\end{center}
\end{figure}

For strong magnetic fields, variation of radiative opacity with $B$ can be modelled as $\kappa = \kappa_B \approx 5.5\times10^{31}\rho T^{-1.5} B^{-2} {\rm cm^{2}\ g^{-1}}$ \cite{PY2001,VP2001}. The field dependent Potekhin's opacity is used instead of Kramers' opacity if $B/10^{12}\ {\rm G} \geq T/10^6\ {\rm K}$, which is valid for the strong $B$ cases that we consider here. The left panel of Figure \ref{fig5} shows the effect of luminosity on the mass-radius relation for non-magnetized WDs compared to Chandrasekhar's results \cite{Ch35}. Although the increase in $L$ leads to progressively higher masses for larger WDs, Chandrasekhar mass limit is retained irrespective of the luminosity. The right panel of Figure \ref{fig5} shows the effect of magnetic field on the mass-radius relation for B-WDs with $L=10^{-4}\, L_{\odot}$ and compares them with the non-magnetic Chandrasekhar results. It can be seen that the magnetic field affects the mass-radius relation in a manner analogous to increasing $L$, by shifting the curve towards higher masses for WDs with larger radii. The mass-radius curves for $B_0 \lesssim 10^{13}\, {\rm G}$ practically overlap with each other and retain the Chandrasekhar mass limit. However, for strong central fields with $B_0 \sim 10^{14}\, {\rm G}$, super-Chandrasekhar WDs are obtained with masses as high as $\sim 1.9\, M_{\odot}$. 

In order to ensure structural stability of a B-WD, an increase in magnetic energy density has to be compensated by a corresponding decrease in the thermal energy and hence the luminosity. This effect is especially prominent for B-WDs with larger radii where the magnetic, thermal and gravitational energies are comparable with each other. We find that a slight decrease in the luminosity for $R \gtrsim 12000\, {\rm km}$ WDs leads to masses that are similar to their non-magnetic counterparts. However, the smaller radii B-WDs require a substantial drop in their luminosity (well outside the observable range) and still do not really achieve masses that are similar to the non-magnetized WDs.

\section{Effect of cooling evolution and field dissipation}
Magnetic fields inside a WD undergo decay by Ohmic dissipation and Hall drift processes with timescales given by \cite{HK98,Cumming02}
\begin{eqnarray}
t_{\rm Ohm} = (7\times10^{10}\ {\rm yr})\, \rho_{c,6}^{1/3} R_{4}^{1/2} (\rho_{\rm avg}/\rho_{\rm c}),\\
t_{\rm Hall} = (5\times10^{10}\, {\rm yr})\ l_8^2 B_{0,14}^{-1} T_{\rm c,7}^{2} \rho_{\rm c,10},
\label{tOhm_tHall} 
\end{eqnarray}
where $\rho_{\rm c,n} = \rho_c/10^n\, {\rm g\, cm^{-3}}$, $R_4 = R/10^4\, {\rm km}$, $T_{c,7}=T_c/10^7\, {\rm K}$, $B_{0,14}=B_0/10^{14}\, {\rm G}$ and $l = l_8 \times 10^8\, {\rm cm}$ is characteristic length scale of the flux loops through WD outer core. Ohmic decay is the dominant field dissipation process for $B \lesssim 10^{12}\ {\rm G}$, while for $10^{12} \leq B/{\rm G} \leq 10^{14}$ the decay occurs via Hall drift and for $B \gtrsim 10^{14}\ {\rm G}$, the principal decay mechanism is likely to be ambipolar diffusion \cite{HK98}. The magnetic field decay in magnetars with surface fields between $10^{14}$ and $10^{16}\, {\rm G}$ can be solved using
\begin{equation}
\frac{{\rm d}B}{{\rm d}t} = -B\left(\frac{1}{t_{\rm Ohm}} + \frac{1}{t_{\rm Amb}} + \frac{1}{t_{\rm Hall}}\right),
\end{equation}
where $t_{\rm Amb}$ denotes the ambipolar diffusion time scale. 

On the other hand, the thermal energy of WDs is radiated away gradually over time in the observed luminosity from the surface layers as the star evolves. The rate at which thermal energy of ions is transported to surface and radiated depends on the specific heat, given by 
\begin{equation}
L = -\frac{d}{dt}\int c_{v} dT = (2\times10^6\ {\rm erg/s)}\, \frac{Am_{\mu}}{M_{\odot}}\left(\frac{T}{K}\right)^{7/2},
\end{equation}
where $c_v \approx 3k_B/2$ is the specific heat at constant volume. Given an initial $L$ and temperature $T_0$ at time $t_0$, final temperature after cooling is given by $(T/{\rm K})^{-5/2} - (T_0/{\rm K})^{-5/2} = 2.406\times10^{-34}\, \tau/{\rm s}$, where $\tau = t-t_0$ is the WD age.

\begin{figure}[h]%
\begin{center}
  \parbox{2.25in}{\includegraphics[width=2.25in]{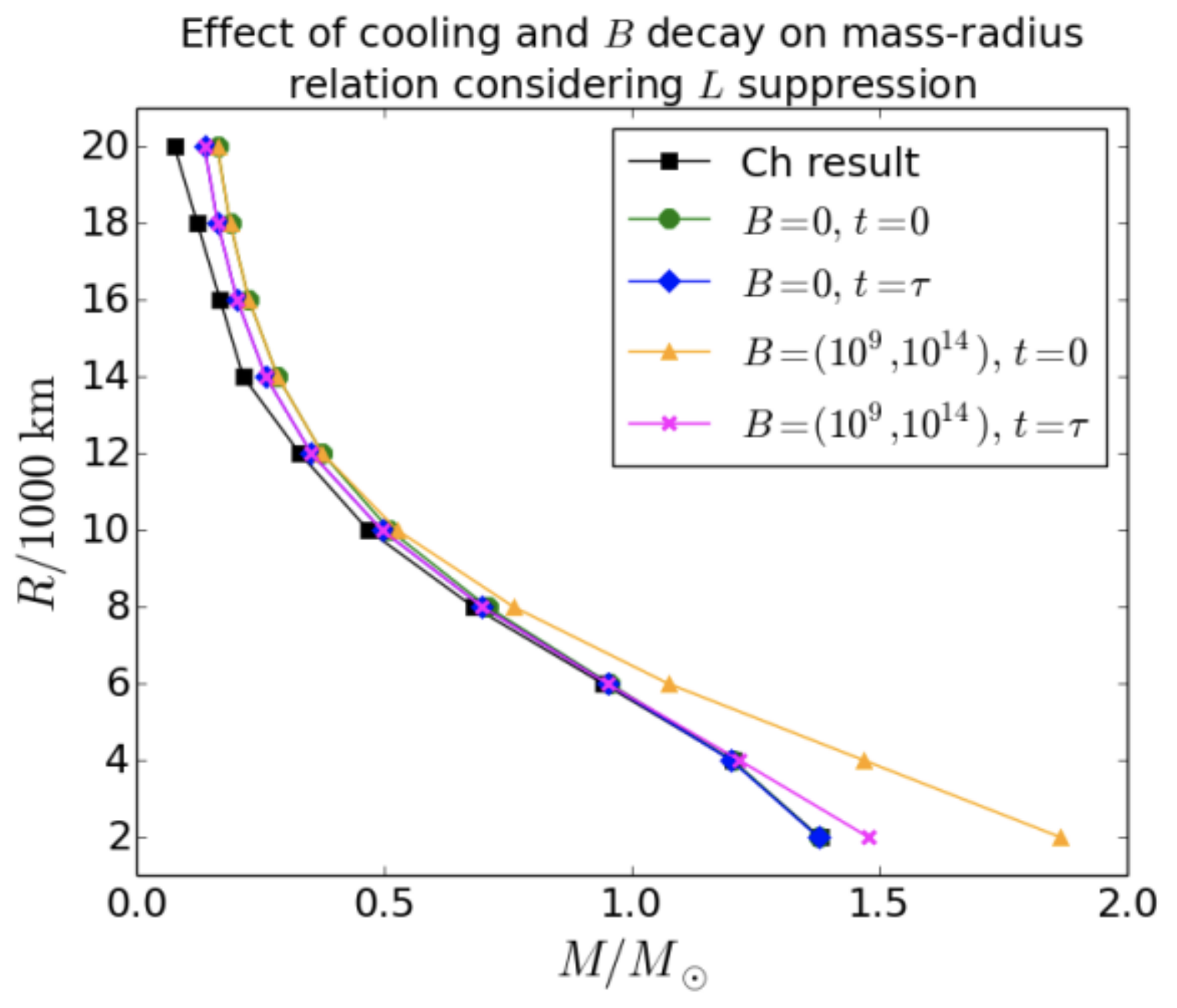}}
  \hspace*{1.0pt}
  \parbox{2.6in}{\includegraphics[width=2.6in]{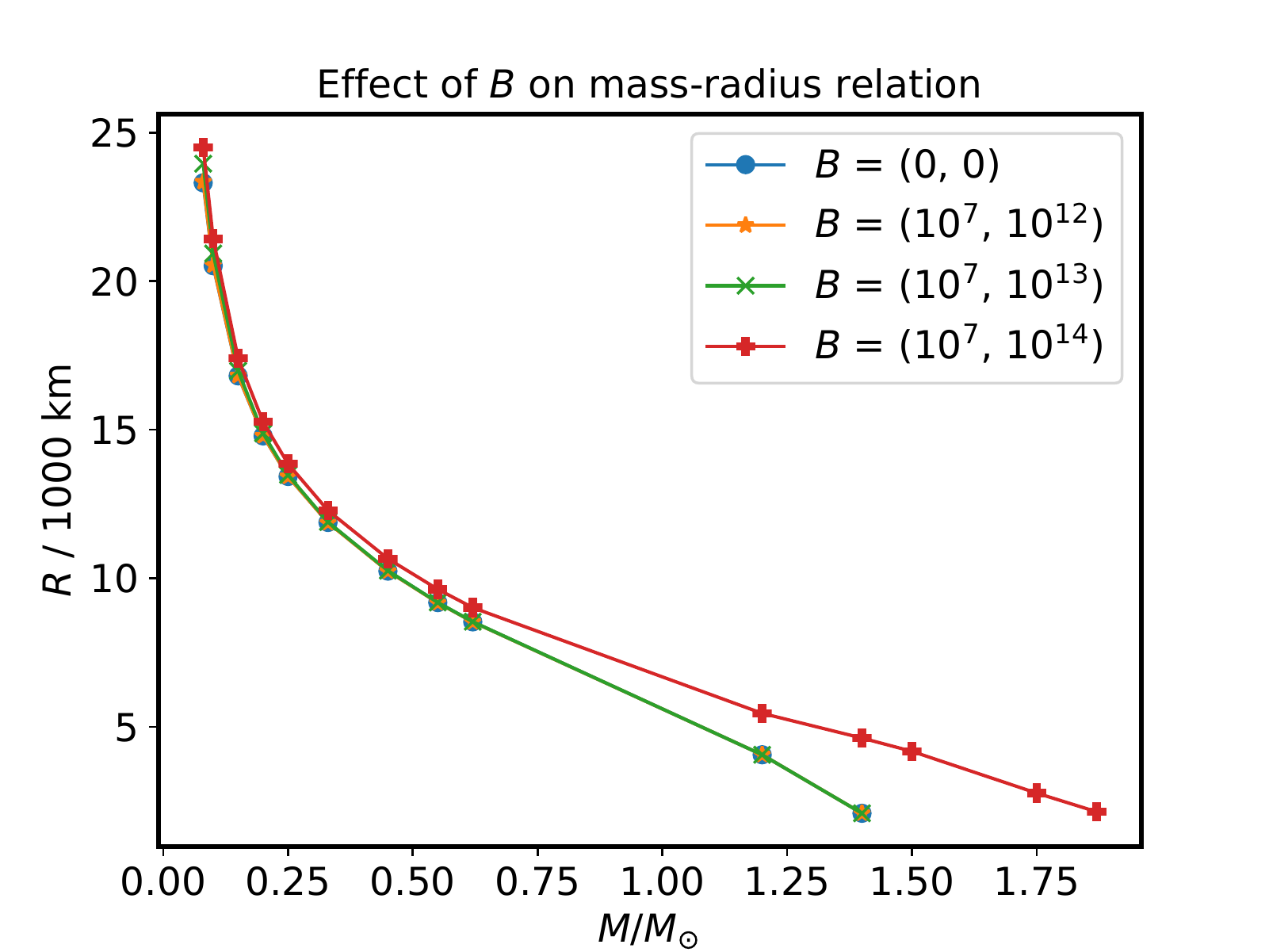}}
  \caption{\emph{Left panel:} The effect of magnetic field on the WD luminosity set to match with the non-magnetised mass--radius relation for the analytical model. The results are shown for $B=(0,0)$ at initial time $t=0$ (green circles), $B=(0,0)$ at $t=10\ {\rm Gyr}$ (blue diamonds), $B=(10^9,10^{14})\, {\rm G}$ at $t=0$ (orange triangles) and $B=(10^9,10^{14})\, {\rm G}$ at $t=10\ {\rm Gyr}$ (magenta crosses). \emph{Right panel:} STARS results to show the effect of magnetic field on the mass--radius relation of highly magnetized WDs for $B=(0, 0)$ (blue circles), $B=(10^7, 10^{12})\, {\rm G}$ (orange stars), $B=(10^7, 10^{13})\, {\rm G}$ (green crosses) and $B=(10^7, 10^{14})\, {\rm G}$ (red pluses).}%
\label{fig6}
\end{center}
\end{figure}

The left panel of Figure \ref{fig6} shows the effect of B-WD evolution on their mass-radius relations including both magnetic field decay and thermal cooling effects. The luminosities are varied with field strength such that the masses can match those obtained for the non-magnetized WDs.  For $B=(B_s,B_0)=(10^9,10^{14})\, {\rm G}$, although the maximum mass $\sim 1.9\, M_{\odot}$ shown in the track at small radius turns out to be much larger than the Chandrasekhar limit, we find that it is lowered considerably to $\sim 1.5\, M_{\odot}$ primarily as a result of magnetic field decay and also thermal cooling over $t=10\, {\rm Gyr}$.

Further, we use the STARS stellar evolution code to qualitatively investigate the B-WD mass-radius relationship at different field strengths, with the objective of numerically validating our semi-analytical models. In the right panel of Figure \ref{fig6}, we show the mass-radius relations obtained from STARS. We obtain results that are in good agreement with our analytical formalism and the magnitude of $B_0$ dictates the shape of the mass-radius curve. In validation of our analytical approach, we have found that the limiting mass $\sim 1.8703\, M_{\odot}$ obtained with the STARS numerical models is in very good agreement with $M \approx 1.87\, M_{\odot}$ that is inferred from the semi-analytical calculations for WDs with strong fields $B=(10^{6-9},10^{14})\, {\rm G}$. We argue that the young super-Chandrasekhar B-WDs only sustain their large masses up to $\sim 10^{5}-10^6\, {\rm yr}$, and this essentially explains their apparent scarcity even without the difficulty of detection owing to their suppressed luminosities.

\section{Continuous gravitational waves from magnetized white dwarfs}
\begin{figure}[h]%
\begin{center}
  \parbox{2.1in}{\includegraphics[width=2.1in]{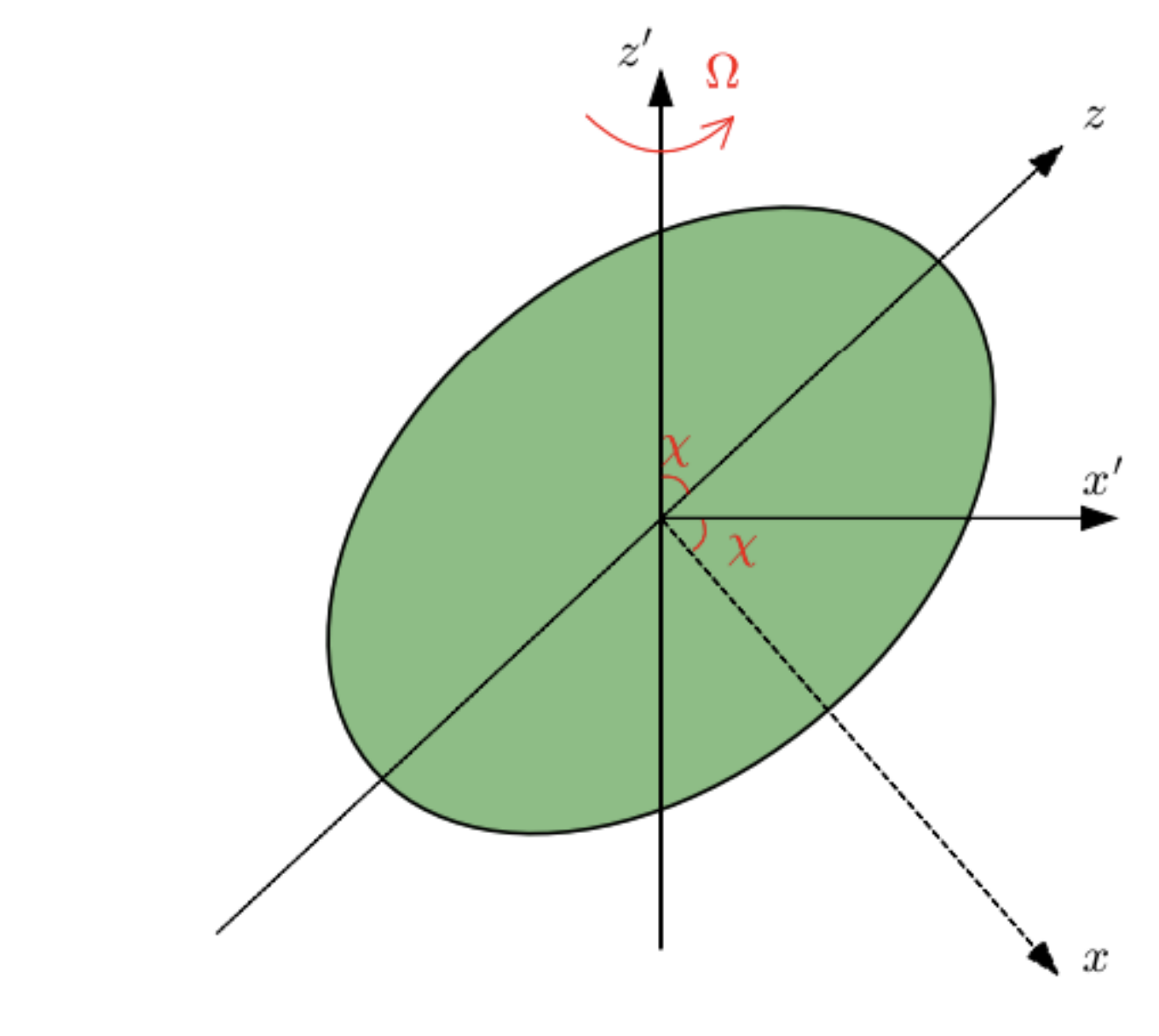}}
  \hspace*{1.0pt}
  \parbox{2.8in}{\includegraphics[width=2.8in]{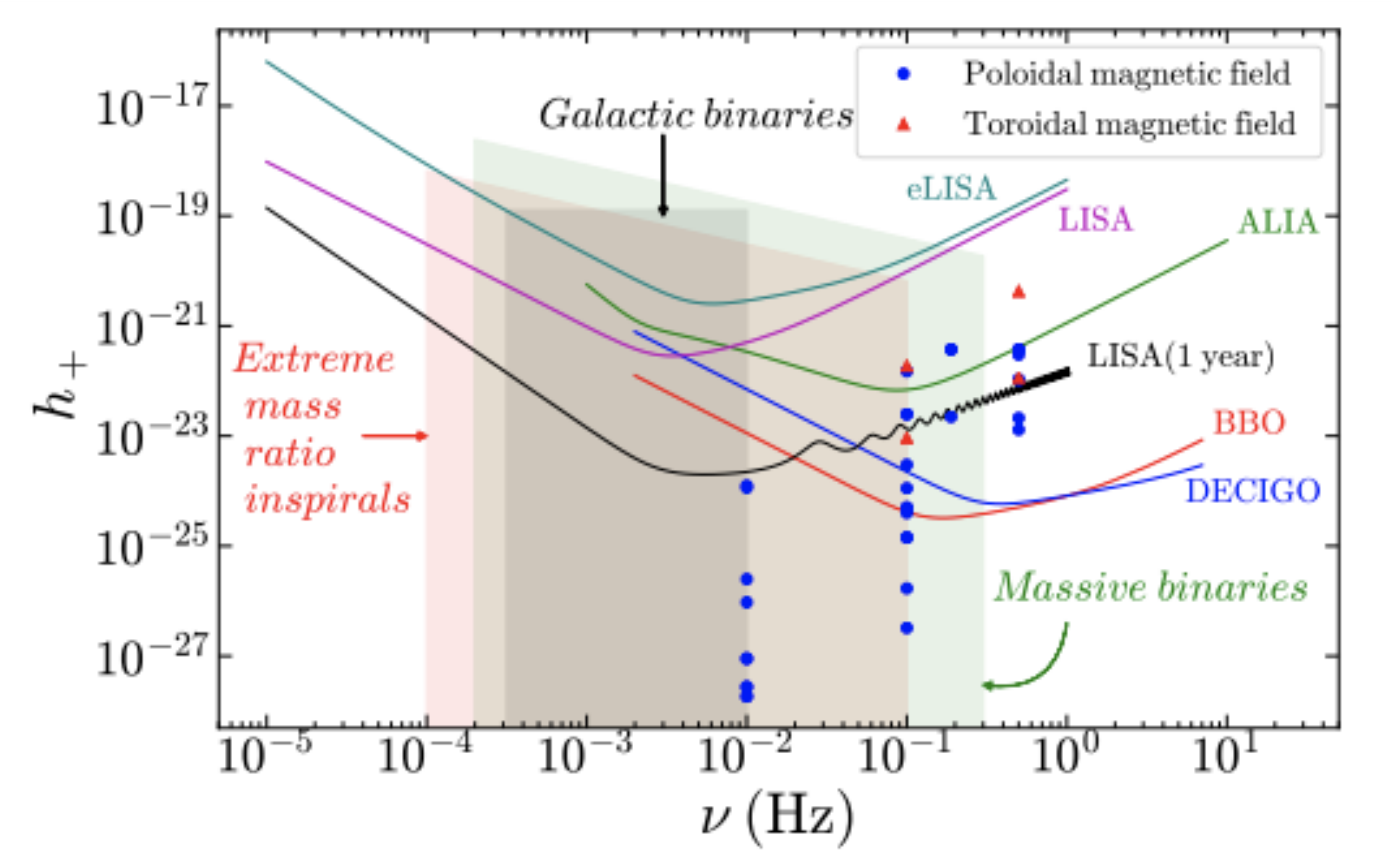}}
  \caption{\emph{Left panel:} Schematic diagram of a B-WD with $z^{\prime}$ being the rotational axis and $z$ the magnetic axis. \emph{Right panel:} The dimensionless GW amplitudes for WDs are shown as functions of frequency, along with the sensitivity curves of various detectors. Optimum $i$ is chosen for $\chi$ at $t=0$.}%
\label{fig7}
\end{center}
\end{figure}

The question then remains, how to detect B-WDs directly.
Continuous gravitational waves can be among the alternate ways to detect super-Chandrasekhar WD candidates directly. If these B-WDs are rotating with certain angular frequency, then they can efficiently emit gravitational radiation, provided that their magnetic field and rotation axes are not aligned \cite{BG96}, and these gravitational waves can be detected by upcoming instruments such as LISA, BBO, DECIGO, Einstein Telescope, etc. The dimensionless amplitudes of the two polarizations of the gravitational wave (GW) at a time $t$ are given by \cite{BG96,ZS79}
\begin{eqnarray}
h_+ = h_0 {\rm sin}\,\chi \, \left[\frac{1}{2}{\rm cos}\,i\, {\rm sin}\,i \,{\rm cos}\,\chi \, {\rm cos}\,\Omega t - \frac{1+{\rm cos}^2 i}{2}{{\rm sin}\,\chi\,}{\rm cos}\,2\Omega t\right], \nonumber \\
h_{\times} = h_0 {\rm sin}\,\chi \left[\frac{1}{2}{\rm sin}\,i\, {\rm cos}\,\chi\, {\rm sin}\,\Omega t - {\rm cos}\,i\, {\rm sin}\,\chi\, {\rm sin}\,2\Omega t \right],
\end{eqnarray}
with $h_0 = (-6G/c^4)Q_{z^{\prime}z^{\prime}}(\Omega^2/d)$, where $Q_{z^{\prime}z^{\prime}}$ is the quadrupole moment of the distorted star, $\chi$ is the angle between the rotation axis $z^{\prime}$ and the body's third principal axis $z$, $i$ is the angle between the rotation axis of the object and our line of sight. The left panel of Figure \ref{fig7} shows a schematic diagram of a pulsar with $z^{\prime}$ being the rotational axis and $z$ the magnetic field axis, where the angle between these two axes is $\chi$. The GW amplitude is
\begin{equation}
h_0 = \frac{4G}{c^4} \frac{\Omega^2 \epsilon I_{xx}}{d},
\end{equation}
where $\epsilon = (I_{zz} - I_{xx})/I_{xx}$ is the ellipticity of the body and $I_{xx}$, $I_{yy}$, $I_{zz}$ are the principal moments of inertia. Note, we have used the XNS code \cite{Pili14} to simulate the underlying axisymmetric equilibrium configuration of B-WDs in general relativity. Moreover, we assume the distance between the WD and the detector to be 100 pc.

Since a pulsating WD can emit both dipole and gravitational radiations simultaneously, it is associated with both dipole and quadrupolar luminosities. The dipole luminosity for an axisymmetric WD is given by \cite{Melatos00}
\begin{equation}
L_D = \frac{B_p^2 R_p^6 \Omega^4}{2c^3} {\rm sin}^2 \, \chi \, F(x_0),
\end{equation}
where $x_0 = R_0 \Omega/c$, $B_p$ is the magnetic field strength at the pole, $R_p$ is radius of the pole and $R_0$ is the average WD radius. The function $F(x_0)$ is defined as 
\begin{equation}
F(x_0) = \frac{x_0^4}{5(x_0^6 - 3x_0^4 + 36)} + \frac{1}{3(x_0^2 + 1)}.
\end{equation}
Similarly, the quadrupolar GW luminosity is given by \cite{ZS79} 
\begin{equation}
L_{GW} = \frac{2G}{5c^5}(I_{zz} - I_{xx})^2 \Omega^6 {\rm sin}^2\, \chi \, (1+15{\rm sin}^2 \, \chi).
\end{equation}
It should be noted that this formula is valid if $\chi$ is very small. The total luminosity is due to both dipole and gravitational radiations. Therefore, the changes in $\Omega$ and $\chi$ with time are dependent on both $L_D$ and $L_{GW}$. The variations of $\Omega$ and $\chi$ with respect to time are given by \cite{Melatos00}
\begin{eqnarray}
\nonumber
\frac{d(\Omega I_{z^{\prime}z^{\prime}})}{dt} = -\frac{2G}{5c^5}(I_{zz}-I_{xx})^2 \Omega^5 {\rm sin}^2\, \chi \, (1+15\, {\rm sin}^2\, \chi) - \frac{B_p^2 R_p^6 \Omega^3}{2c^3}{\rm sin}^2\, \chi \, F(x_0), \\
\\
\nonumber
I_{z^{\prime}z^{\prime}}\frac{d\chi}{dt} = -\frac{12G}{5c^5}(I_{zz} - I_{xx})^2 \Omega^4 {\rm sin}^3\, \chi\, {\rm cos}\, \chi - \frac{B_p^2 R_p^6 \Omega^2}{2c^3}{\rm sin}\, \chi \, {\rm cos}\, \chi \, F(x_0),\\
\end{eqnarray}
where $I_{z^{\prime}z^{\prime}}$ is the moment of inertia about $z^{\prime}$-axis. Equations (17) and (18) need to be solved simultaneously to obtain the timescale over which a WD can radiate. 

The right panel of Figure \ref{fig7} shows the dimensionless GW amplitudes for the WDs as functions of their frequencies, along with the sensitivity curves of various detectors. It can be seen that the isolated WDs may not be detected directly by LISA, but can be detected after integrating the signal to noise ratio $S/N$ for 1 year. As WDs are larger in size compared to NS, they cannot rotate as fast as NS and hence ground-based GW detectors such as LIGO, Virgo and KAGRA are not expected to detect the isolated WDs. 
These isolated WDs are also free from the noise due to the galactic binaries as well as from the extreme mass ratio inspirals (EMRIs).

\section{Summary \& Conclusions}
We have shown that highly magnetized, stable WDs, namely B-WDs, have a variety of implications, including enigmatic peculiar over-luminous SNe Ia. Numerical simulations utilizing the stellar evolution code STARS indicate that the central field in strongly magnetized B-WDs can be toroidally dominated whereas the surface fields are more of dipole nature. The 
mass $\sim 1.87\, M_{\odot}$ obtained for these B-WDs from the STARS numerical models, as of now, is totally consistent with the estimates from the analytical stellar structure models for strong fields $B= (10^{6-9}, 10^{14})\, {\rm G}$. These young super-Chandrasekhar B-WDs have suppressed luminosities and are difficult to observe or rare, due to their decaying magnetic fields. We have found that these WDs may not remain super-Chandrasekhar for long i.e. beyond $\sim 10^{5-6}\, {\rm year}$ due to decaying field primarily, and indicate rapidly losing pulsar nature and/or low luminosity. However, these stars can be very promising candidates for GW detectors such as LISA (with 1 year integrated $S/N$) and also for Einstein Telescope and future DECIGO/BBO missions. Therefore, appropriate missions in GW astronomy and otherwise, e.g. radio astronomy, should be planned in order to probe them in the future.

\section*{Acknowledgments}
B.M. acknowledges a partial support
by a project of Department of Science and Technology (DST-SERB) with research Grant No. DSTO/PPH/BMP/1946
(EMR/2017/001226).
M.B. acknowledges support from Graduate Continuing Fellowship at the University of Texas, Austin. A.J.H. thanks
the Science and Technology Facilities Council (STFC) and
the Cambridge Commonwealth, European \& International
Trust for his doctoral funding. C.A.T. thanks Churchill College for his fellowship.




\end{document}